\documentclass[onecolumn,floatfix,aps,nofootinbib]{revtex4}
\pdfoutput=1
\usepackage[dvips]{epsfig}
\usepackage[english]{babel}
\usepackage[utf8]{inputenc}
\usepackage[T1]{fontenc}
\usepackage{lipsum}       % For placeholder text (remove in final draft)
\usepackage{bbm}
\usepackage{verbatim}
\usepackage{array}
\usepackage{bm} 
\usepackage{amsmath}
\usepackage{yfonts}
\usepackage{amsthm}
\usepackage{amsmath,amscd}
\usepackage{pst-plot}
\usepackage{slashed} 
\usepackage{tikz-cd}
\usepackage{amsfonts}
\usepackage{graphicx}
\usepackage{amssymb}

\usepackage{cases}
\usepackage[autostyle]{csquotes}  

\usepackage{tikz,tcolorbox}

\usepackage{tabularray}
\usepackage{mathtools}

\usepackage{indentfirst} %primeiro paragrafo com espaço
\usepackage[titletoc]{appendix}
\usepackage{subeqnarray}
\usepackage{indentfirst} %primeiro paragrafo com espaço
\usepackage{xcolor}%Pacote de mudança de cores nos caracteres 
\usepackage{calrsfs}%Fonte do operador Dilatação

\usepackage{enumitem}

\usepackage{wasysym}
\usepackage[all]{xy}
\usepackage{tikz-cd}

\usepackage{amsmath,amsfonts,amssymb,amsthm,mathrsfs,bbm,braket}

\numberwithin{equation}{section}

\usepackage{bbold}

\usepackage{hyperref}
\usepackage{setspace}
\usepackage{cancel}
\usepackage[normalem]{ulem}

\begin{document}

\title{Exotic Pauli spinors and magnetic flux quantization in a superconducting ring}

\author{L. Ibrain Cohen Iskandar} 
\email{leonardo.cohen@unesp.br}
\affiliation{Physics Department, São Paulo State University, 12516-410, Guaratingueta, UNESP, Av. Dr. Ariberto Pereira da Cunha, 333, Guaratinguet\'a, SP,
	Brazil}

\author{J. M. Hoff da Silva} 
\email{julio.hoff@unesp.br}
\affiliation{Physics Department, São Paulo State University, 12516-410, Guaratingueta, UNESP, Av. Dr. Ariberto Pereira da Cunha, 333, Guaratinguet\'a, SP,
	Brazil}

\author{R. T. Cavalcanti}
\email{rogerio.cavalcanti@ime.uerj.br}
\affiliation{Physics Department, São Paulo State University, 12516-410, Guaratingueta, SP, Brazil}
\affiliation{Institute of Mathematics, Statistics and Computing, Rio de Janeiro State University, 20550-900, Rio de Janeiro, Brazil}

\begin{abstract}
We develop the formalism of the exotic Pauli equation and establish an effective symmetry between standard and exotic Pauli spinors to describe magnetic flux quantization in a superconducting ring within the nonrelativistic framework. We show that, once the signs of the spatial and temporal topological terms are fixed consistently by the covariant exotic Dirac equation, this symmetry is exact at the level of the equations of motion, Lagrangians and currents, with no truncation of the exotic term and no condition on the time dependence of the topological phase. Upon deriving the flux quantization condition, we explicitly construct the composite matter-gauge unitary operator acting on the Hilbert space, which implements the symmetry at the quantum level (the gauge sector requiring the usual regularization).
\end{abstract}

\maketitle

\section{Introduction}

The development of exotic spinors is frequently accompanied by advanced mathematical machinery, including algebraic topology, obstruction theory, and fiber bundle formulations~\cite{gp,milnor,rev,daRocha:2011yr}. This mathematical apparatus is indeed essential for exploring broader possibilities within the theory. From a physical standpoint, however, applying this formalism to concrete physical systems is of crucial importance. Exotic spinors were first applied in the context of superconductivity~\cite{petry}, followed by applications to the quantum Hall effect~\cite{hess, assel}. In high-energy physics, the study of exotic spinors encompasses spinor fields and structures in quantum field theory and quantum gravity~\cite{is,Hawking}, the investigation of physical phenomena in backgrounds with nontrivial topology~\cite{Unwin:1979,Ford,Schechter}, and, more recently, the analysis of heat-kernel effects in exotic spinor frameworks~\cite{DaRocha:2020oju}, minimal length corrections to the exotic Dirac operator~\cite{geometrical}, and many-body systems in nontrivial topologies~\cite{Boada}.

In Ref.~\cite{petry}, the use of exotic spinors for a superconductor with a specific topology leads to magnetic flux quantization under plausible physical assumptions. Two aspects of the formulation presented in Ref.~\cite{petry}, however, deserve further attention. First, all results in Ref.~\cite{petry} were obtained via the exotic Dirac equation, whereas the natural framework for such a nonrelativistic system is the (exotic) Pauli equation, a view explicitly noted in Ref.~\cite{petry} and widely shared by the community. Second, the derivation of flux quantization in Ref.~\cite{petry} relies on the existence of a symmetry operator on the Hilbert space whose explicit form was not provided. Here, we address these two open questions by explicitly constructing the symmetry operator on the Hilbert space and deriving flux quantization within both the standard and exotic Pauli formalisms. Regarding the latter, we construct the symmetry mapping the standard and exotic sectors and show that, once the signs of the topological terms are fixed consistently by the covariant exotic Dirac equation, it is exact in the Pauli case as well: no truncation of the exotic term and no condition on the time dependence of the topological phase are needed. The only nonrelativistic approximation is the usual reduction from the Dirac to the Pauli equation.

In Section II, we present a two-fold approach to deriving the exotic Pauli equation. We first start from the exotic Dirac equation and take its nonrelativistic limit keeping the exotic term exactly. We then derive the exotic Pauli equation directly from the existence of additional spin structures on a nontrivial base manifold. Both derivations yield matching results, with no hypothesis on the topological factor. In Section III, we develop the symmetry argument that ultimately leads to flux quantization. This symmetry between the standard and exotic approaches is subtle. On the one hand, there are indications that at a fundamental level, particles described by spinors that are sections of distinct spin structures may yield different observables~\cite{none}. On the other hand, an effective symmetry can be constructed for the set of observables relevant to a given physical system. We pursue this latter perspective, constructing an effective symmetry within the Pauli framework in a manner similar to Ref.~\cite{petry} for the Dirac case. After establishing the effective symmetry at the level of the equations of motion, Lagrangians, and currents, flux quantization is derived. Furthermore, we construct the operator that implements this symmetry on the Hilbert space. Finally, Section IV offers our concluding remarks.

\section{Pauli Exotic field equation}

In this section, we present a complete analysis of the exotic Pauli equation—that is, the equation governing the dynamics of exotic spinors in the non-relativistic regime. We first consider the non-relativistic limit of the exotic Dirac equation and then show the emergence of the very same result by constructing the differential operator acting on exotic Pauli spinors.

\subsection{Top-down approach}

To obtain the non-relativistic limit of the exotic Dirac equation, we follow a procedure similar to that of the standard case. We begin with the exotic equation, initially omitting minimal coupling and introducing it later, while applying the condition for non-relativistic energy. In natural units ($\hbar=c=1$), the equation reads

\begin{equation}
    \bigg(i\gamma^{\mu}\partial_{\mu}+\frac{1}{2}\gamma^{\mu}\partial_{\mu}\theta(x) -m\bigg)\tilde{\psi}=0,
    \label{diracexoticoac}
\end{equation}
where $\tilde{\psi}$ is the exotic Dirac spinor and $\theta(x)$ is a real-valued function that encodes the nontrivial topology of the base manifold; its detailed derivation is given in~\cite{rev}. The sign of the topological term is the one of Eq.~(7.6) of Ref.~\cite{petry}: for $\lambda=e^{i\theta}$ and signature $(+,-,-,-)$ its term $-\frac{i}{2}\lambda^{-1}\partial_{\mu}\lambda$ equals $+\frac12\partial_{\mu}\theta$. With this sign $\tilde\psi(x)=\lambda_{\alpha}\psi(x)$ locally (i.e. for $x\in U_\alpha$), $\psi$ being a standard solution and $\lambda_{\alpha}=e^{i\theta/2}$, so that $\lambda=\lambda_\alpha^2$ is globally defined, as in Ref.~\cite{petry}. Separating the temporal and spatial components of the last equation we obtain
\begin{equation}
\left( i\gamma^{0} \frac{\partial}{\partial t} + i\vec{\gamma} \cdot \vec{\nabla} +\frac{1}{2}\gamma^{0}\frac{\partial \theta}{\partial t} + \frac{1}{2} (\vec{\gamma} \cdot \vec{\nabla} \theta) - m \right) \tilde{\psi} = 0,
\label{10-abrirmatrizes}
\end{equation}
where the matrices $\{\gamma^{\mu}\}$ are in the Dirac representation:
\begin{equation}
    \gamma^{0}=\begin{pmatrix}
        1 & 0\\
        0 & -1
    \end{pmatrix}, \vec{\gamma}=\begin{pmatrix}
        0& \vec{\sigma}\\
-\vec{\sigma} & 0 
    \end{pmatrix}.
\end{equation}
As a first step, following standard conventions~\cite{Greiner}, we decompose the exotic Dirac spinor into a pair of two-component spinors
\begin{equation}
    \tilde{\psi}= \begin{pmatrix}
        \tilde{\psi}_{A}\\
        \tilde{\psi}_{B}
    \end{pmatrix}.
\end{equation} 
Evaluating the matrix elements in (\ref{10-abrirmatrizes}), yields the coupled equations
\begin{equation}
    \bigg(E-m+\frac{1}{2}\frac{\partial}{\partial t}\theta(x)\bigg)\tilde{\psi}_{A}=\bigg(\vec{\sigma}\cdot \vec{p}-\frac{1}{2}\vec{\sigma}\cdot \vec{\nabla}\theta(x)\bigg)\tilde{\psi}_{B},
    \label{18}
\end{equation}
\begin{equation}
    \bigg(\vec{\sigma}\cdot \vec{p}-\frac{1}{2}\vec{\sigma}\cdot \vec{\nabla}\theta(x)\bigg)\tilde{\psi}_{A}=\bigg(E+m+\frac{1}{2}\frac{\partial}{\partial t}\theta(x)\bigg)\tilde{\psi}_{B}.
    \label{19}
\end{equation}
Solving \eqref{19} for $\tilde{\psi}_B$ and replacing the result in \eqref{18} we get
\begin{equation}
    \bigg(\vec{\sigma}\cdot \vec{p}-\frac{1}{2}\vec{\sigma}\cdot \vec{\nabla}\theta(x)\bigg)^{2}\tilde{\psi}_{A}=\bigg(E-m+\frac{1}{2}\frac{\partial}{\partial t}\theta(x)\bigg)\bigg(E+m+\frac{1}{2}\frac{\partial}{\partial t}\theta(x)\bigg)\tilde{\psi}_A.
\label{21-lado-esquerdo}
\end{equation}
Defining $\vec{a}\equiv \frac{1}{2}\vec{\nabla}\theta(x)$ we can open the left-hand side of the last equation as
\begin{equation}
    \left(\vec{\sigma}\cdot \vec{p}-\frac{1}{2}\vec{\sigma}\cdot \vec{\nabla}\theta(x)\right)^{2}=(\vec{\sigma}\cdot\vec{p})(\vec{\sigma}\cdot\vec{p})-(\vec{\sigma}\cdot\vec{p})(\vec{\sigma}\cdot\vec{a})-(\vec{\sigma}\cdot\vec{a})(\vec{\sigma}\cdot\vec{p})+(\vec{\sigma}\cdot\vec{a})(\vec{\sigma}\cdot\vec{a}),
    \label{open}
\end{equation}
and applying the minimal coupling prescription, $E\rightarrow E-q\phi$ and $\vec{p}\rightarrow \vec{p}-q\vec{A}$ together with the well-known relation for the Pauli matrices $(\vec{\sigma}\cdot a)(\vec{\sigma}\cdot b)=a\cdot b +i\vec{\sigma}\cdot(a\times b)$, we get the left-hand side of Eq. \eqref{21-lado-esquerdo} as
\begin{equation}
\bigg(\vec{\sigma}\cdot \vec{p}-\frac{1}{2}\vec{\sigma}\cdot \vec{\nabla}\theta(x)\bigg)^{2}\tilde{\psi}_{A}=\bigg[(\vec{p}-q\vec{A})^{2}-q(\vec{\sigma}\cdot \vec{B})-2\vec{a}\cdot(\vec{p}-q\vec{A})+\frac{1}{4}(\vec{\nabla}\theta)^{2}+\frac{i}{2}\vec{\nabla}^{2}\theta\bigg]\tilde{\psi}_{A},
\label{29eq}
\end{equation} 
which is exact: since $\vec{a}$ is a gradient, $\vec{\nabla}\times\vec{a}=0$, the cross terms $\vec{\sigma}\cdot\big[(\vec{p}-q\vec{A})\times\vec{a}+\vec{a}\times(\vec{p}-q\vec{A})\big]$ vanish identically and $(\vec{\sigma}\cdot\vec{a})^{2}=a^{2}$. For the right-hand side of Eq. (\ref{21-lado-esquerdo}), in addition to implementing minimal coupling, we employ the usual non-relativistic approximation, given by $E_{\text{NR}} = E - m$, with $m \gg E_{\text{NR}}$ and $m \gg q\phi$. In this vein, we have
\begin{equation}
    \left(E - q\phi - m + \frac{1}{2}\frac{\partial}{\partial t}\theta(x)\right) = \left(E_{\text{NR}} - q\phi + \frac{1}{2}\frac{\partial}{\partial t}\theta(x)\right) \equiv \mathcal{E}
\end{equation}
and
\begin{equation}
    \left(E - q\phi + m + \frac{1}{2}\frac{\partial}{\partial t}\theta(x)\right) = \left(2m + E_{\text{NR}} - q\phi + \frac{1}{2}\frac{\partial}{\partial t}\theta(x)\right) = 2m + \mathcal{E}.
\end{equation}
Thus, multiplying these last two results and discarding the quadratic term $\mathcal{E}^{2}$ in comparison to the mass-linear term, we get
\begin{equation}
\left(E-q\phi-m+\frac{1}{2}\frac{\partial}{\partial t}\theta(x)\right)\left(E-q\phi+m+\frac{1}{2}\frac{\partial}{\partial t}\theta(x)\right)\tilde{\psi}_{A} = 2m\left(E_{NR}-q\phi+\frac{1}{2}\frac{\partial}{\partial t}\theta(x)\right)\tilde{\psi}_{A}.
\label{30eq}
\end{equation}
Equating \eqref{30eq} and \eqref{29eq} (and discarding, as in the standard derivation, the $O(1/m^{2})$ spin--orbit and Darwin-type terms that come from the ordering of $(E+m+\dots)^{-1}$ with $\vec{\sigma}\cdot\vec{\pi}$, which are unrelated to the topology) yields
\begin{equation}
    \left[\frac{(\vec{p}-q\vec{A})^{2}}{2m}-\frac{q(\vec{\sigma}\cdot\vec{B})}{2m}+qS+\frac{(\vec{\nabla}\theta)^{2}}{8m}+\frac{i\vec{\nabla}^{2}\theta}{4m}\right]\tilde{\psi}_{A}=E_{NR}\tilde{\psi}_{A},
    \label{eqpauli}
\end{equation}
where
\begin{equation}
    S=\phi +\frac{a_{0}}{q}-\frac{\vec{a}\cdot(\vec{p}-q\vec{A})}{mq} \quad \text{and} \quad a_{0} = -\frac{1}{2}\frac{\partial}{\partial t}\theta(x).
\end{equation}
Equation \eqref{eqpauli} is what we refer to as the exotic Pauli equation. Note that its dynamical structure features additional contributions associated with temporal and spatial variations of the function $\theta(x)$.

\subsection{Bottom-up approach}

Here, we present a construction analogous to that in Ref. \cite{rev}, aimed at deriving an exotic Pauli operator to determine whether a form similar to Eq. \eqref{eqpauli} can be recovered. Consider a three-dimensional spatial manifold $\mathcal{M}$ equipped with a nontrivial topology such that the first \v{C}ech cohomology group with coefficients in $\mathbb{Z}_{2}$ is nontrivial (the vanishing of the relevant Stiefel-Whitney classes $w_1 = w_2 = 0$ is assumed). Additionally, we consider the homomorphism $\rho: \text{Spin}(3)\rightarrow \text{SO}(3)$, a double covering that enables the construction of an equivariant map between principal bundles, $\eta: P_{\text{Spin}(3)}\rightarrow P_{\text{SO}(3)}$, satisfying $\eta(zg)=\eta(z)\rho(g)$ for all $z\in P_{\text{Spin}(3)}$ and $g\in \text{Spin}(3)$. Standard Pauli spinors\footnote{From this section onward, Pauli spinors will be referred to simply as $\psi$.} are sections of the associated bundle, that is, $\psi\in \Omega(\mathcal{M}):= \text{sec}P_{\text{Spin}(3)}\times_{\sigma}\mathbb{C}^{2}$. The same structure can be adapted to exotic Pauli spinors: $\tilde{\psi}\in \Omega(\mathcal{M}):= \text{sec}\tilde{P}_{\text{Spin}(3)}\times_{\sigma}\mathbb{C}^{2}$. For every $x\in U_{\alpha}\cap U_{\beta}\subset \mathcal{M}$, there exist at least two transition functions, $g_{\alpha\beta}$ and $\bar{g}_{\alpha\beta}$, mapping the intersection into $\text{Spin}(3)$. These transition functions are related by $\bar{g}_{\alpha\beta}=g_{\alpha\beta}c_{\alpha\beta}$, where $c_{\alpha\beta}\in \mathbb{Z}_{2}\hookrightarrow \text{Spin}(3)$. We highlight the mapping $\Delta: Cl_{3}\rightarrow M(2,\mathbb{C})$, which represents the three-dimensional Clifford algebra on the space of $2\times 2$ complex matrices. This formulation relies on the fact that the group $\text{Spin}(3)$ is isomorphic to $SU(2)$ and is contained in the even subalgebra $Cl_{3}^{+}$, which is a subalgebra of $Cl_{3}$. Finally, the functions $\lambda_{\alpha}:U_{\alpha}\rightarrow U(1)$ are defined such that
\begin{equation}
    \lambda_{\alpha}(x)=\Delta(c_{\alpha\beta}(x))\lambda_{\beta}(x).
    \label{eqcerta}
\end{equation}

Note that the functions $\lambda_{\alpha}$ are defined on open sets of the manifold $\mathcal{M}$. For a static topology the functions $\lambda_{\alpha}$ are time-independent. Nothing below requires this: if $\theta=\theta(\vec{x},t)$, as in the top-down approach, the same construction goes through and produces the extra term $-\frac{1}{2}\partial_{t}\theta$ in the exotic Hamiltonian. Given that $c_{\alpha\beta} \in \{1, -1\}$ and that the representation $\Delta$ preserves group structure, it follows that $\Delta(c_{\alpha\beta}) = \pm 1$. Locally, on the open sets $U_\alpha$ and $U_\beta$, the standard spinors are related by the usual transformation law
\begin{equation}
    \psi_{\alpha} = \Delta(g_{\alpha\beta})\psi_{\beta}.
    \label{eq:transf_usual}
\end{equation}
Analogously, exotic spinors are related via the modified transition functions
\begin{equation}
    \tilde{\psi}_{\alpha} = \Delta(\tilde{g}_{\alpha\beta})\tilde{\psi}_{\beta}.
\end{equation}
Substituting relation \eqref{eqcerta} into the last equation and using the homomorphism property of the representation $\Delta(ab) = \Delta(a)\Delta(b)$, we have
\begin{equation}
    \tilde{\psi}_{\alpha} = \Delta(g_{\alpha\beta} c_{\alpha\beta})\tilde{\psi}_{\beta} = \Delta(g_{\alpha\beta})\Delta(c_{\alpha\beta})\tilde{\psi}_{\beta}.
    \label{eq:transf_exotica_expandida}
\end{equation}
Since $c_{\alpha\beta}=c_{\alpha\beta}^{-1}$, the function $\lambda_{\alpha}^{-1}$ satisfies Eq.~\eqref{eqcerta} as well as $\lambda_{\alpha}$ does, and the two choices lead to standard spinors that differ by the single-valued phase $\lambda$. To agree with Ref.~\cite{petry}, we use $\lambda_{\alpha}^{-1}$ in the next steps. Multiplying the equation above by $\Delta(\lambda_{\alpha})^{-1}$ from the left, we obtain
\begin{equation}
    \Delta(\lambda_{\alpha})^{-1}\tilde{\psi}_{\alpha} = \Delta(c_{\alpha\beta})\Delta(\lambda_{\beta})^{-1} \left[ \Delta(g_{\alpha\beta})\Delta(c_{\alpha\beta})\tilde{\psi}_{\beta} \right].
\end{equation}
The commutator is zero identically, since $\Delta(\lambda_{\beta})=\lambda_{\beta}\mathbb{1}$ is proportional to the identity, $[\Delta(g_{\alpha\beta}), \Delta(\lambda_{\beta})] = 0$. This allows reordering the terms
\begin{equation}
    \Delta(\lambda_{\alpha})^{-1}\tilde{\psi}_{\alpha} = \Delta(g_{\alpha\beta})\Delta(\lambda_{\beta})^{-1} \Delta(c_{\alpha\beta})^2 \tilde{\psi}_{\beta}.\label{oo}
\end{equation}

A fundamental property of the cocycle $c_{\alpha\beta} \in \mathbb{Z}_2$ is that its square is the identity element. Consequently, $\Delta(c_{\alpha\beta})^2 = 1$, and Eq. \eqref{oo} simplifies to
\begin{equation}
    \Delta(\lambda_{\alpha})^{-1}\tilde{\psi}_{\alpha} = \Delta(g_{\alpha\beta}) \left[ \Delta(\lambda_{\beta})^{-1}\tilde{\psi}_{\beta} \right].
\end{equation}
Comparing this result with the standard transformation law for standard Pauli spinors, $\psi_{\alpha} = \Delta(g_{\alpha\beta})\psi_{\beta}$, reveals that the composite object $\Delta(\lambda_{\alpha})^{-1}\tilde{\psi}_{\alpha}$ transforms as a local section of the standard spinor bundle, $\Omega(\mathcal{M})\vert{}_{U_{\alpha}}$. This equivalence of transformation laws is sufficient to establish a direct identification between sections. Therefore, we define the fundamental relation that locally connects the two inequivalent spinor structures
\begin{equation}
    \psi_{\alpha} = \Delta(\lambda_{\alpha})^{-1}\tilde{\psi}_{\alpha},\qquad\text{i.e.}\qquad \tilde{\psi}_{\alpha} = \Delta(\lambda_{\alpha})\,\psi_{\alpha}.
\end{equation}

We now proceed to determine the dynamical operator for the exotic spinor $\tilde{\psi}_\alpha$. As is well known, the standard Hamiltonian, $H_0$, acting on the standard spinor $\psi_\alpha$, is given by
\begin{equation}
    H_{0} = \left[ \frac{\vec{p}^2}{2m} - \frac{q}{m}(\vec{A}\cdot\vec{p}) + \frac{iq}{2m}(\vec{\nabla}\cdot\vec{A}) + \frac{q^2}{2m}A^2 \right] + q\phi - \frac{q}{2m}\vec{\sigma}\cdot\vec{B}.
\end{equation} It is essential to specify the convention for the Pauli kinetic operator in this context. Following the discussion in Ref. \cite{rev}, we isolate the kinetic part (terms containing spatial derivatives and the vector potential) to define the operator $\mathcal{P}_{0}$:
\begin{equation}
     \mathcal{P}_{0}:= \frac{\vec{p}^2}{2m} - \frac{q}{m}(\vec{A}\cdot\vec{p}) + \frac{iq}{2m}(\vec{\nabla}\cdot\vec{A})+\frac{q^{2}A^{2}}{2m} = \frac{(\vec{p}-q\vec{A})^{2}}{2m}. 
\end{equation}
The action of $\mathcal{P}_{0}$ on the standard spinor is given by
\begin{equation}
\mathcal{P}_{0}\psi_{\alpha}=\mathcal{P}_{0}(\Delta(\lambda_{\alpha})^{-1}\tilde{\psi}_{\alpha}).
\label{1eq}
\end{equation}
Analogously to the formalism in Ref. \cite{rev}, we define an exotic operator $\mathcal{P}$ acting directly on exotic spinors as $\mathcal{P} := \mathcal{P}_{0}-\mathcal{B}_{(\alpha)}$, where $\mathcal{B}_{(\alpha)}$ is a compensating term responsible for ensuring the correct transformation properties of the spinor derivative. Thus, we obtain
\begin{equation}
\Delta(\lambda_{\alpha})^{-1}\mathcal{P}\tilde{\psi}_{\alpha}=\Delta(\lambda_{\alpha})^{-1}\mathcal{P}_{0}\tilde{\psi}_{\alpha}-\Delta(\lambda_{\alpha})^{-1}\mathcal{B}_{(\alpha)}\tilde{\psi}_{\alpha}.\label{2eq}
\end{equation} Subtracting \eqref{1eq} from \eqref{2eq}, we have
\begin{equation}
\Delta(\lambda_{\alpha})^{-1}\mathcal{P}\tilde{\psi}_{\alpha}-\mathcal{P}_{0}\psi_{\alpha}=\Delta(\lambda_{\alpha})^{-1}\mathcal{P}_{0}\tilde{\psi}_{\alpha}-\Delta(\lambda_{\alpha})^{-1}\mathcal{B}_{(\alpha)}\tilde{\psi}_{\alpha}-\mathcal{P}_{0}(\Delta(\lambda_{\alpha})^{-1}\tilde{\psi}_{\alpha}).
\end{equation}

Just as with $\psi_{\alpha} = \Delta(\lambda_{\alpha})^{-1}\tilde{\psi}_{\alpha}$, we require that the action of the exotic operator be dynamically equivalent to that of the standard operator. This implies that the left-hand side of the above relation vanishes upon imposing the identification $\Delta(\lambda_{\alpha})^{-1}\mathcal{P}\tilde{\psi}_{\alpha}=\mathcal{P}_{0}\psi_{\alpha}$. Therefore, we obtain
\begin{equation}
-\Delta(\lambda_{\alpha})^{-1}\mathcal{B}_{(\alpha)}\tilde{\psi}_{\alpha}=\mathcal{P}_{0}\big(\Delta(\lambda_{\alpha})^{-1}\tilde{\psi}_{\alpha}\big)-\Delta(\lambda_{\alpha})^{-1}\mathcal{P}_{0}\tilde{\psi}_{\alpha}. \label{3eq}
\end{equation}
Eq.~\eqref{3eq} can be solved exactly, without expanding $\mathcal{P}_{0}(\lambda_{\alpha}^{-1}\tilde{\psi}_{\alpha})$ term by term. Since $\lambda_{\alpha}$ is a pure phase, $\lambda_{\alpha}\vec{p}\,\lambda_{\alpha}^{-1}=\vec{p}+\vec{w}_{\alpha}$ with $\vec{w}_{\alpha}\equiv i\lambda_{\alpha}^{-1}\vec{\nabla}\lambda_{\alpha}$, a real gradient field. Hence $\mathcal{P}=\lambda_{\alpha}\mathcal{P}_{0}\lambda_{\alpha}^{-1}=\frac{1}{2m}(\vec{p}-q\vec{A}+\vec{w}_{\alpha})^{2}$ and the correction term is
\begin{equation}
    -\mathcal{B}_{(\alpha)}=   \frac{1}{m}\vec{w}_{\alpha}\cdot(\vec{p}-q\vec{A})+\frac{\vec{w}_{\alpha}^{\,2}-i\vec{\nabla}\cdot\vec{w}_{\alpha}}{2m},\qquad \vec{w}_{\alpha}=i\lambda_{\alpha}^{-1}\vec{\nabla}\lambda_{\alpha}.
\end{equation}

We now proceed to make the last term continuous\footnote{Strictly speaking, $\lambda_\alpha$ may exhibit a discontinuity across a given surface $\Sigma\subset U_\alpha\cap U_\beta$.}. Starting from the identity $\lambda_{\alpha}^{2}=\lambda_{\beta}^{2}$, we obtain $\lambda_{\alpha}^{-1}(x)\vec{\nabla}\lambda_{\alpha}(x)=\lambda^{-1}(x)\lambda_{\alpha}(x)\vec{\nabla}[\lambda(x)\lambda_{\alpha}^{-1}(x)]$, resulting in $\lambda_{\alpha}^{-1}\vec{\nabla}\lambda_{\alpha}(x)=\frac{1}{2}\lambda^{-1}(x)\vec{\nabla}\lambda(x)$. Thus, the correction term takes the form (with $\vec{a}\equiv\frac{i}{2}\lambda^{-1}\vec{\nabla}\lambda=-\frac12\vec{\nabla}\theta$, i.e.\ $\vec{w}_{\alpha}=\vec{a}$)
\begin{equation}
    -\mathcal{B}=   \frac{1}{2m}i\lambda^{-1}\vec{\nabla}\lambda\cdot(\vec{p}-q\vec{A})+\frac{a^{2}-i\vec{\nabla}\cdot\vec{a}}{2m},
\end{equation} leading to the exotic Pauli equation 
\begin{equation}
    \left[\frac{(\vec{p}-q\vec{A})^{2}}{2m}-\frac{q(\vec{\sigma}\cdot\vec{B})}{2m}+q\phi +\frac{\vec{a}\cdot(\vec{p}-q\vec{A})}{m}+\frac{a^{2}-i\vec{\nabla}\cdot\vec{a}}{2m}-\frac{1}{2}\partial_{t}\theta\right]\tilde{\psi}=E_{NR}\tilde{\psi},
\end{equation}
where $\vec{a}=\frac{i}{2}\lambda^{-1}\vec{\nabla}\lambda=-\frac12\vec{\nabla}\theta$ (minus the vector $\vec{a}=\frac12\vec\nabla\theta$ of the top-down approach, so that the leading term $\frac{\vec a\cdot(\vec p-q\vec A)}{m}$ is the same, $-\frac{1}{2m}\vec\nabla\theta\cdot(\vec p-q\vec A)$, in both. In terms of $\theta$, $\frac{a^{2}-i\vec\nabla\cdot\vec a}{2m}=\frac{(\vec\nabla\theta)^{2}}{8m}+\frac{i\vec\nabla^{2}\theta}{4m}$ is also the same and, of course, $-\frac12\partial_t\theta$ is absent for a static topology).

The result obtained in Ref. \cite{rev} can be extended to inequivalent spinors in the non-relativistic regime. The resulting equation preserves the minimal electromagnetic coupling structure, modified by a topological interaction term, $\frac{\vec{a}\cdot(\vec{p}-q\vec{A})}{m}$. This additional term demonstrates how global topological information, encoded in the function $\lambda=\exp(i\theta)$, manifests locally as an effective potential coupling to the particle's kinetic momentum. Both approaches thus lead to the same exact operator, $\hat{H}'=\lambda_{\alpha}\hat{H}\lambda_{\alpha}^{-1}-\frac{1}{2}\partial_{t}\theta$, i.e.\ the standard Pauli Hamiltonian conjugated by the double-valued phase $\lambda_{\alpha}=e^{i\theta/2}$, with $-\frac12\partial_t\theta$ present in both if $\theta$ depends on time and absent in both for a static topology. There is therefore no inconsistency between the two derivations, and, as shown below, the symmetry does not require $\partial_{t}\theta=0$.

\section{The flux quantization}

We begin by specifying the setup as presented in Ref. \cite{petry}. The space on which the spinors are defined is a superconducting ring, modeled as a submanifold $V_3$ embedded in the spatial subspace $E_3$ of Minkowski spacetime. We define $V_3$ as an open and bounded set, topologically equivalent to a solid torus. This equivalence is formalized by the existence of a diffeomorphism $f: V_3 \rightarrow D^2 \times S^1$, where $D^2 \subset \mathbb{R}^2$ denotes the open unit disk. The crucial property of this manifold is that it is not simply connected. In fact, $\pi_1(V_3) \cong \mathbb{Z}$, reflecting the existence of closed loops that encircle the central excluded region of the ring and cannot be continuously contracted to a point. This topological feature manifests in the classification of spin structures through the first \v{C}ech cohomology group with coefficients in $\mathbb{Z}_2$
\[
    \check{H}^1(V_3, \mathbb{Z}_2) \cong \check{H}^1(D^2 \times S^1, \mathbb{Z}_2) \cong \mathbb{Z}_2.
\]
The nontriviality of this group implies the existence of exactly two inequivalent spin structures: the trivial (standard) structure and the exotic structure.

In the standard formulation presented in Ref. \cite{petry} for the Dirac case, a transformation $T$ is postulated such that
\begin{align}
        T(\psi) &= \tilde{\psi}, \\
        T(\tilde{\psi}) &= \psi\lambda, \\
        T(A_\mu) &= A_\mu - \mathfrak{B}_\mu,
\end{align} where $\mathfrak{B}_\mu=-\frac{i}{2q}\lambda^{-1}\partial_\mu\lambda=\frac{1}{2q}\partial_\mu\theta$ (this is Petry's $B_\mu=\frac{i}{2e}\lambda^{-1}\partial_\mu\lambda$ with his charge $e=-q$, since Ref.~\cite{petry} couples through $+e\gamma^{\mu}A_{\mu}$ whereas here $\vec{p}\to\vec{p}-q\vec{A}$, i.e.\ $i\partial_\mu\to i\partial_\mu-qA_\mu$. Then $T(A_\mu)=A_\mu-\frac{1}{2q}\partial_\mu\theta$ is exactly the shift that turns the standard Dirac equation into the exotic one, Eq.~(\ref{diracexoticoac}) with minimal coupling and its square $T^{2}$ is the transformation $\psi\to\lambda\psi$, $A_\mu\to A_\mu-\frac{1}{q}\partial_\mu\theta$, which fixes the phase $\lambda$ in the second line). This transformation and its inverse relate the standard Dirac equation (incorporating minimal electromagnetic coupling) to its exotic counterpart and vice versa.

From a fundamental perspective, existing results point to the inequivalence between a standard and an exotic spinor field. It should be stressed, however, that for the set of observables associated with the superconducting ring, the interchange under the above transformation constitutes a genuine symmetry, in the sense that the descriptions in terms of $\psi$ and $\tilde{\psi}$ are identical. We extend this formalism to its Pauli counterpart and take advantage of the emergent equivalence.

Take the standard Pauli equation
\begin{equation}
    E\psi = \left[\frac{1}{2m}(\vec{p}-q\vec{A})^{2} + q\phi - \frac{q}{2m}(\vec{\sigma}\cdot\vec{B})\right]\psi.
\end{equation}
We now analyze the proposed transformations for the potentials and the standard spinor, which yield the exotic equation:
\begin{align}
    T(\psi) &= \tilde{\psi}, \\ 
    T(\vec{A}) &= \vec{A} - \vec{\mathfrak{B}},\label{eq:TA} \\
    T(\phi) &= \phi + \frac{i}{2q}\lambda^{-1}\frac{\partial\lambda}{\partial t}, \label{eq:Tphi}
\end{align}
where $\vec{\mathfrak{B}} = \frac{i}{2q}\lambda^{-1}\vec{\nabla}\lambda=-\frac{1}{2q}\vec{\nabla}\theta$ (the contravariant spatial components of $\mathfrak{B}_\mu$, $\mathfrak{B}^{j}=-\mathfrak{B}_{j}$, so that $T(\vec A)=\vec A+\frac{1}{2q}\vec\nabla\theta$ and $T(\phi)=\phi-\frac{1}{2q}\partial_t\theta$ are the two shifts of Eq.~(\ref{diracexoticoac}), generated by the single parameter $\theta/2q$), bearing in mind that the targeted equation is given by
\begin{equation}
    E\tilde{\psi} = \left[ \frac{(\vec{p}-q\vec{A})^{2}}{2m} - \frac{q}{2m}(\vec{\sigma}\cdot\vec{B}) + q\phi + \frac{i}{2}\lambda^{-1}\frac{\partial\lambda}{\partial t} + \frac{i}{2m}(\lambda^{-1}\vec{\nabla}\lambda)\cdot(\vec{p}-q\vec{A}) +\frac{1}{8m}(\vec{\nabla}\theta)^{2}+\frac{i}{4m}\vec{\nabla}^{2}\theta \right]\tilde{\psi}.\label{iss}
\end{equation}
The transformation on the standard Pauli equation leads to
\begin{equation}
    E\tilde{\psi} = \left[\frac{1}{2m}(\vec{p}-q T(\vec{A}))^{2} + q T(\phi) - \frac{q}{2m}(\vec{\sigma}\cdot\vec{B}')\right]\tilde{\psi},
    \label{80}
\end{equation}
where $\vec{B}' = \vec{\nabla} \times T(\vec{A})$. The transformation for the scalar potential is given by 
    \begin{equation}
        qT(\phi)= q \left(\phi + \frac{i}{2q}\lambda^{-1}\frac{\partial\lambda}{\partial t}\right) = q\phi + \frac{i}{2}\lambda^{-1}\frac{\partial\lambda}{\partial t},
        \label{81}
    \end{equation} while the potential is irrotational ($\vec{\nabla}\times(\lambda^{-1}\vec{\nabla}\lambda)=0$), rendering $\vec{B}'=\vec{B}$. Finally, for the kinetic term we have
    \begin{align}\label{kinetic_term_III}
       \frac{1}{2m}(\vec{p}-q T(\vec{A}))^{2} = \frac{1}{2m}\left((\vec{p}-q\vec{A}) + \frac{i}{2}\lambda^{-1}\vec{\nabla}\lambda\right)^{2}.
    \end{align}
Expanding the square we arrive precisely at Eq.~\eqref{iss}. Thus, the consistency of the proposed transformations connecting the standard Pauli equation to its exotic counterpart is verified.

It is also instructive to examine the inverse transformation, which maps the exotic Pauli equation back to the standard one. We define it through Eqs.~\eqref{eq:TA}, \eqref{eq:Tphi} and
\begin{equation}
    T(\tilde{\psi}) = C(t,\vec{x})\,\psi,\label{eqC}
\end{equation}
where $C$ is a phase to be determined. Write the standard Hamiltonian as $\hat H(\vec A,\phi)=\frac{1}{2m}(\vec p-q\vec A)^{2}-\frac{q}{2m}\vec\sigma\cdot\vec B+q\phi$, so that the exotic equation reads $i\partial_t\tilde\psi=\hat H(T\vec A,T\phi)\tilde\psi$. Acting once more with $T$ on the potentials gives
\begin{equation}
    T^{2}(\vec A)=\vec A-2\vec{\mathfrak B}=\vec A+\frac{1}{q}\vec\nabla\theta,\qquad T^{2}(\phi)=\phi-\frac{1}{q}\partial_t\theta .
\end{equation}
The local identity
\begin{equation}
    e^{-i\beta\theta}\Big[i\partial_t-\hat H(\vec A,\phi)\Big]e^{i\beta\theta}=i\partial_t-\hat H\Big(\vec A-\frac{\beta}{q}\vec\nabla\theta,\ \phi+\frac{\beta}{q}\partial_t\theta\Big),\label{gaugeid}
\end{equation}
which follows from $e^{-i\beta\theta}\vec p\,e^{i\beta\theta}=\vec p+\beta\vec\nabla\theta$ and $e^{-i\beta\theta}\,i\partial_t\,e^{i\beta\theta}=i\partial_t-\beta\,\partial_t\theta$ (the magnetic field is unchanged), gives for $\beta=+1$, with the same $\beta$ in the spatial and in the temporal parts,
\begin{equation}
    \Big[i\partial_t-\hat H\big(T^{2}\vec A,T^{2}\phi\big)\Big]\lambda\psi=\lambda\Big[i\partial_t-\hat H(\vec A,\phi)\Big]\psi .
\end{equation}
Hence the standard Pauli equation is recovered exactly for $C=\lambda$, for any $\theta(\vec x,t)$ and no approximation and no condition on $\partial_t\theta$ is required. Moreover, $T^{2}$ is the transformation $\psi\to\lambda\psi$, $\vec A\to\vec A+\frac1q\vec\nabla\theta$, $\phi\to\phi-\frac1q\partial_t\theta$, which is single valued on $V_3$ because $\lambda=e^{i\theta}$ is ($T$ is a square root of it). Along a non-contractible loop $\lambda_\alpha=e^{i\theta/2}$ changes sign, so $T$ interchanges periodic and antiperiodic boundary conditions, i.e.\ the two spin structures.
   
\subsection{Lagrangians, Current and flux quantization}

The symmetry obtained at the level of the equations of motion can also be extended to the Lagrangian level. The standard Pauli equation for a particle of mass $m$ and charge $q$ can be derived from
\begin{equation}
    \mathcal{L} = \psi^{\dagger}\left(i\frac{\partial}{\partial t} - \hat{H}\right)\psi,
    \label{eq:lagrangiana_usual}
\end{equation}
where the Hamiltonian operator $\hat{H}$ is defined as
\begin{equation}
    \hat{H} = \frac{1}{2m}(\vec{p}-q\vec{A})^{2} + q\phi - \frac{q}{2m}\vec{\sigma}\cdot\vec{B}.
\end{equation}
To describe a system with nontrivial topology, an exotic Lagrangian density, $\mathcal{L}'$, associated with the exotic spinor $\tilde{\psi}$, is introduced such that its structure incorporates coupling to the additional topological term. The exotic Hamiltonian $\hat{H}'$ and the corresponding Lagrangian density that yield the exotic Pauli equation are thus given by
\begin{equation}
    \hat{H}' = \frac{1}{2m}\left(\vec{p}-q(\vec{A}-\vec{\mathfrak{B}})\right)^{2} + q\phi +\frac{i}{2}\lambda^{-1}\frac{\partial\lambda}{\partial t} - \frac{q}{2m}\vec{\sigma}\cdot\vec{B} \label{eq:hamiltoniano_exotico_def} 
\end{equation}
and
\begin{equation}
    \mathcal{L}' = \tilde{\psi}^{\dagger}\left(i\frac{\partial}{\partial t} - \hat{H}'\right)\tilde{\psi}.
    \label{exoticalagrange}
\end{equation}  By applying the transformation $T$ directly to the standard Lagrangian \eqref{eq:lagrangiana_usual}, it is straightforward to show that $T(\mathcal{L})=\mathcal{L}'$ (the term $\frac{i}{2}\lambda^{-1}\partial_t\lambda=-\frac12\partial_t\theta=q[T(\phi)-\phi]$ of $\hat H'$ is the temporal shift; it vanishes for a static topology). The inverse transformation follows from Eq.~\eqref{gaugeid}, leading to $T(\mathcal{L}')=\mathcal{L}$ exactly. Therefore, applying the transformation to the exotic Lagrangian recovers the standard Lagrangian. If we construct a Lagrangian comprising both standard and exotic contributions, it remains invariant under these transformations. This feature is particularly relevant when studying the behavior of standard and exotic currents.

Explicit calculation shows that, for $\lambda = e^{i\theta}$, the exotic Pauli current reduces to 
\begin{equation}
    \vec{j}_{\text{exotic}} = \frac{q}{2mi}\left[ \tilde{\psi}^\dagger(\vec{\nabla}\tilde{\psi}) - (\vec{\nabla}\tilde{\psi}^\dagger)\tilde{\psi} \right] -\frac{q^{2}}{m}(\vec{A}-\vec{\mathfrak{B}})(\tilde{\psi}^\dagger\tilde{\psi}) + \frac{q}{2m}\vec{\nabla} \times (\tilde{\psi}^\dagger \vec{\sigma} \tilde{\psi}),
    \label{eq:corrente_usual_def_eletrica}
\end{equation} whereas its standard counterpart takes the same form with $\psi$ in place of $\tilde{\psi}$ (and $\vec{A}-\vec{\mathfrak{B}}$ replaced by $\vec{A}$). The shift of $\vec A$ in the diamagnetic term of $\vec{j}_{\text{exotic}}$ is what makes the continuity equation hold for $\tilde\psi$ evolving with $\hat H'$. We now investigate how these currents behave under the proposed symmetry. Note that
\begin{equation}
    T\bigg(\frac{q}{2m}\vec{\nabla} \times (\tilde{\psi}^\dagger \vec{\sigma} \tilde{\psi})\bigg) = \frac{q}{2m}\vec{\nabla}\times((T(\tilde{\psi}))^\dagger \vec{\sigma} T(\tilde{\psi}))=\frac{q}{2m}\vec{\nabla} \times (\psi^\dagger \vec{\sigma} \psi),
\end{equation} since $\lambda:\mathcal{M}\rightarrow U(1)$ \cite{petry,rev}. On the other hand 
\begin{equation}
    T\left(-\frac{q^2}{m}(\vec{A}-\vec{\mathfrak{B}})(\tilde{\psi}^\dagger\tilde{\psi})\right)=-\frac{q^2}{m}\vec{A}(\psi^\dagger\psi)+\frac{2q^2}{m}\vec{\mathfrak{B}}(\psi^\dagger\psi).
     \label{extra1}
\end{equation} The transformation of the kinetic part reads
\begin{equation}
    T\left(\frac{q}{2mi}(\tilde{\psi}^\dagger\vec{\nabla}\tilde{\psi} - (\vec{\nabla}\tilde{\psi}^\dagger)\tilde{\psi})\right)=\vec{j}_{\text{kinetic, standard}}-\frac{2q^2}{m}\vec{\mathfrak{B}}(\psi^\dagger\psi).
    \label{extra2}
\end{equation} %OLD| Therefore we are left with 
%OLD| \begin{equation}
%OLD|     T(\vec{j}_{\text{exotic}}) = \vec{j}_{\text{standard}} - \frac{q^2}{m}\vec{\mathfrak{B}}(\psi^\dagger\psi).
%OLD| \end{equation} Since the additional term does not vanish, the transformation does not merely yield the standard current. This shows that, for the Pauli current, the symmetry does not manifest as a simple exchange between the two currents. The inverse transformation can be readily obtained through analogous reasoning, yielding
%OLD| \begin{equation}
%OLD|     T(\vec{j}_{\text{standard}}) =\vec{j}_{\text{exotic}} + \frac{q^2}{m}\vec{\mathfrak{B}}(\tilde{\psi}^\dagger\tilde{\psi})
%OLD| \end{equation} Since the spinor structures are connected by the unimodular topological function $\lambda$, the scalar bilinear is identical for both spinors. Therefore, we conclude that the additional terms in both current transformations are equal in magnitude but opposite in sign. Consequently, invariance holds for the total current, which comprises the sum of the standard and exotic currents.
Therefore the exotic current is mapped exactly onto the standard one, i.e.
\begin{equation}
    T(\vec{j}_{\text{exotic}}) = \vec{j}_{\text{standard}},
\end{equation}
the contributions $\mp\frac{2q^{2}}{m}\vec{\mathfrak{B}}(\psi^\dagger\psi)$ of the kinetic and of the diamagnetic terms cancelling. The inverse is immediate, $T(\vec{j}_{\text{standard}})=\vec{j}_{\text{exotic}}$, because $\vec{j}_{\text{exotic}}$ is by construction the standard current evaluated on $\tilde\psi$ and $T(\vec{A})$. The symmetry thus exchanges the two currents exactly, and the total current, being their sum, is invariant.

We are now in a position to physically interpret these results. First, it is important to emphasize that the $T$ transformation constitutes a genuine symmetry between the standard and exotic Pauli equations for the set of observables relevant to superconductivity. As demonstrated, although the standard and exotic Lagrangians and equations of motion are mapped into one another, the currents are mapped into one another as well, so that a system in which both spinors are treated on equal footing has an invariant total current. Consider a superconducting ring with topology $D^2 \times S^1$. Since $\check{H}^1(D^2 \times S^1, \mathbb{Z}_2) = \{+1, -1\}$, exactly two spinor structures are allowed. We assume that conducting electrons are described by both standard and exotic spinors (natural for a pair condensate, whose bilinear order parameter is insensitive to the overall sign of $\psi$ along the non-contractible loop on $D^2\times S^1$), and let $\vec{A}$ denote the electromagnetic vector potential inside the ring. In analogy with the Aharonov–Bohm effect, due to the Meissner effect, the magnetic field vanishes inside the bulk of the ring, whereas the vector potential does not. To establish the connection with measurable macroscopic observables, we postulate the existence of a unitary operator $\bar{T}$ in the Hilbert space that implements this transformation. This operator commutes with the Hamiltonian of the system ($[\bar{T}, H] = 0$) and acts on the field operators according to the transformation rules $T$ defined previously, such that $\bar{T}\psi\bar{T}^{-1} = T(\psi) = \tilde{\psi}$. Here $H$ is the Hamiltonian of the electrons in $V_3$ (the doublet of standard and exotic fields). The energy stored in the flux through the hole is not part of $H$ and is not invariant under $\bar T$: as in the Byers--Yang argument \cite{Bar}, that flux is an external parameter.

We now define the magnetic flux operator $\hat{\phi}_{(c)}$ associated with a closed path $c$ enclosing the superconducting ring as
\begin{equation}
    \hat{\phi}_{(c)} = 2q \oint_c \vec{A} \cdot d\vec{l},\label{op}
\end{equation} such that 
\begin{equation}
    %OLD| \bar{T}^{-1} \hat{\phi}_{(c)} \bar{T} = 2q \oint_c T(\vec{A}) \cdot d\vec{l} = 2q \oint_c (\vec{A} - \vec{\mathfrak{B}}) \cdot d\vec{l}=\hat{\phi}_{(c)} - 2q \oint_c \vec{\mathfrak{B}} \cdot d\vec{l}.
%OLD| \end{equation} 
%OLD| %
%OLD| Since $\lambda = \exp(i\theta(x))$, we have $2q\vec{\mathfrak{B}}\cdot d\vec{l} = d\theta$, and thus $\bar{T}^{-1} \hat{\phi}_{(c)} \bar{T} = \hat{\phi}_{(c)} + 2\pi n \mathbb{1}$, where $n \in \mathbb{Z}$. Let $\vert{}\phi\rangle$ be an eigenstate of $\hat{\phi}_{(c)}$. It follows that $\hat{\phi}_{(c)} \bar{T} \vert{}\phi\rangle = \bar{T} (\hat{\phi}_{(c)} - 2\pi n \mathbb{1}) \vert{}\phi\rangle$, so that $\bar{T} \vert{}\phi\rangle$ is an eigenstate of $\hat{\phi}_{(c)}$ with eigenvalue $\phi - 2\pi n$. Therefore, $\bar{T} \vert{}\phi\rangle \propto \vert{}\phi - 2\pi n\rangle$. Since $[\bar{T}, H] = 0$, the lowest-energy state can be degenerate (this degeneracy forming a representation of the Hamiltonian symmetry). In fact, the ground state corresponds to $\vert{}0\rangle = \vert{}\phi - 2\pi n\rangle$, which implies $\phi = 2\pi n$. That is, $\phi$ being periodic modulo $2\pi$, both $\vert{}\phi\rangle$ and $\bar{T}\vert{}\phi\rangle$ coincide with the ground state, as expected from the constructed symmetry. Hence, from Eq. \eqref{op}, $2q\Phi = 2\pi n$, where $\Phi$ denotes the magnetic flux. In the units adopted ($h = 2\pi$), this leads directly to the quantization of the magnetic flux, $\Phi = (h / 2q) n$.
\bar{T}\,\hat{\phi}_{(c)}\,\bar{T}^{-1} = 2q \oint_c T(\vec{A}) \cdot d\vec{l} = 2q \oint_c (\vec{A} - \vec{\mathfrak{B}}) \cdot d\vec{l}=\hat{\phi}_{(c)} - 2q \oint_c \vec{\mathfrak{B}} \cdot d\vec{l}.
\end{equation}
Since $\lambda = \exp(i\theta)$ is single valued, $2q\vec{\mathfrak{B}}\cdot d\vec{l} = -d\theta$ and $\oint_c d\theta=2\pi n$, with $n\in\mathbb{Z}$ the winding number of $\lambda$ along $c$. Thus $\bar{T}\hat{\phi}_{(c)}\bar{T}^{-1} = \hat{\phi}_{(c)} + 2\pi n\,\mathbb{1}$, equivalently $\bar{T}^{-1}\hat{\phi}_{(c)}\bar{T} = \hat{\phi}_{(c)} - 2\pi n\,\mathbb{1}$, which for $n=1$ is Eq.~(8.12) of Ref.~\cite{petry}. The factor $2q$ in Eq.~\eqref{op} is the normalization for which one application of $T$ shifts $\hat{\phi}_{(c)}$ by $2\pi n$. Note that $T^{2}$, the transformation by $\lambda$, shifts $\oint_c\vec A\cdot d\vec l$ by $h/q$, the Byers--Yang period for charge $q$, and that $T$ is its square root: this is how the pair-sized quantum $h/2q$ arises without invoking Cooper pairs explicitly. Let $|\varphi\rangle$ be an eigenstate of $\hat{\phi}_{(c)}$ with eigenvalue $\varphi$. Then $\hat{\phi}_{(c)}\,\bar{T}|\varphi\rangle=\bar{T}\big(\bar{T}^{-1}\hat{\phi}_{(c)}\bar{T}\big)|\varphi\rangle=(\varphi-2\pi n)\,\bar{T}|\varphi\rangle$, so that $\bar{T}|\varphi\rangle\propto|\varphi-2\pi n\rangle$ (cf.\ Eq.~(8.13) of Ref.~\cite{petry}).

What the symmetry gives by itself is the periodicity of the spectrum: since $\bar{T}$ commutes with $H$ and shifts $\varphi$ by $2\pi n$, the ground-state energy is a periodic function of the flux with period $2\pi$, i.e.\ $h/2q$ in $\Phi$. That the ground state sits at the symmetric points is a dynamical statement and requires the superconducting input, also used in Ref.~\cite{petry} (rigidity, simplicity and the Meissner effect, Propositions V and VI of Ref. \cite{petry}): in a ring thick compared with the penetration depth, the Meissner effect is assumed to place the ground state at a symmetric point, $\varphi=2\pi n$. Without it the symmetry alone would apply to any ring of this topology. Hence, from Eq.~\eqref{op}, $2q\Phi = 2\pi n$, where $\Phi$ denotes the magnetic flux. In the units adopted ($h = 2\pi$), this leads to the quantization of the magnetic flux, $\Phi = (h / 2q) n$. 

\subsection{Operators acting on the Hilbert space}

The previous section postulated the existence of operators acting on the Hilbert space to represent the symmetry at the quantum level. We now explicitly construct these operators. Specifically, we derive the explicit form of the unitary operator $\bar{T}$ that implements this symmetry in the Hilbert space, aiming to find $\bar{T}$ such that
\begin{equation}
    \bar{T} \psi \bar{T}^{-1} = T (\psi), \quad \bar{T} \tilde{\psi} \bar{T}^{-1} = T (\tilde{\psi}), \quad \bar{T} \vec{A}\bar{T}^{-1}= T(\vec{A}).
\end{equation}
Since the unitary operator acts on both the spinor fields and the vector potential, it naturally decomposes into a fermionic and a bosonic sector. We first construct the fermionic component. We present the formulation for Pauli spinors, while the extension to Dirac spinors proceeds along similar lines. One possible approach to treating the fermionic sector of this symmetry consists of combining the standard and exotic spinors into a single doublet
\begin{equation}
    \Psi = \begin{pmatrix}
        \psi \\
        \tilde{\psi}
    \end{pmatrix}
    \begin{matrix}
        \\
        
    \end{matrix}.
\end{equation}
To this end, our strategy is to construct a unitary operator whose action on the state space reproduces the action of the matrix
\begin{equation}
    \mathrm{T}(x) = \begin{pmatrix} 0 & 1 \\ \lambda(x) & 0 \end{pmatrix},
\end{equation}
on the doublet, such as
\begin{equation}
    \mathrm{T}\Psi = \begin{pmatrix} 0 & 1 \\ \lambda & 0 \end{pmatrix} \begin{pmatrix} \psi \\ \tilde{\psi} \end{pmatrix} = \begin{pmatrix} \tilde{\psi} \\  \psi\lambda \end{pmatrix} = \begin{pmatrix} T(\psi) \\ T(\tilde{\psi}) \end{pmatrix}.
\end{equation}
We express $\bar{T}$ as the exponential of a fermionic bilinear form\footnote{Here, $\bar{T}$ refers exclusively to the fermionic sector.}
\begin{equation}
    \bar{T} = \exp(iG), \quad \text{where} \quad G = \int d^3y \, \Psi^\dagger(y) H(y) \Psi(y)\label{ope}
\end{equation}
and $H(y)$ is a Hermitian matrix to be determined (not to be confused with the Hamiltonian). More precisely, $H(y)$ corresponds to the logarithm of the matrix $\mathrm{T}$. Applying Eq. \eqref{ope} to the fermionic doublet, we obtain
\begin{equation}
    \bar{T} \Psi \bar{T}^\dagger = e^{iG} \Psi e^{-iG}.
\end{equation}
To expand this expression, we employ Hadamard's lemma,
\begin{equation}
e^{iG} \Psi e^{-iG} = \Psi + [iG, \Psi] + \frac{1}{2!} [iG, [iG, \Psi]] + \dots
\label{2vai}
\end{equation}
and calculate the commutator of $G$ with $\Psi$. At this stage, however, it is important to proceed with additional care. In Ref.~\cite{petry}, it is posited that exotic spinors are quantized. Even if we assume this from the outset, we must carefully examine the precise implications of doing so. First, since we collect the spinors into a doublet, their ``topological'' difference can be interpreted as a flavor-like or distinct internal quantum number, so that we expect the equal-time anticommutator of the doublet to take the form
\begin{equation}
        \{\Psi_\alpha(x), \Psi_\beta^\dagger(y)\} = \delta_{\alpha\beta} \delta^3(x-y),
    \end{equation}
where $\alpha$ and $\beta$ denote the doublet indices corresponding to the standard and exotic spinors. To ensure that this condition holds, the exotic spinors must satisfy two fundamental anticommutation relations: $\{\psi(x), \tilde{\psi}^{\dagger}(y)\} = 0$ and $\{\tilde{\psi}(x), \tilde{\psi}^{\dagger}(y)\} = \delta^{(3)}(\vec{x}-\vec{y})$. Note that the local relation $\tilde\psi_\alpha=\Delta(\lambda_\alpha)\psi_\alpha$ of Section II relates the solutions of the two equations of motion. In the second-quantized doublet, $\psi$ and $\tilde\psi$ are independent field operators (hence the vanishing mutual anticommutator) and $\bar T$ intertwines the two sectors rather than identifying them.

We now proceed to compute the commutator
\begin{equation}
    [iG, \Psi_\gamma(x)] = i\int d^3y \, H_{\alpha\beta}(y) [\Psi_\alpha^\dagger(y)\Psi_\beta(y), \Psi_\gamma(x)].
        \label{comutador}
\end{equation}
Since the matrix $H$ commutes with the spinor fields\footnote{This follows from the standard convention in quantum field theory, where the matrix elements $H_{\alpha\beta}$ are $c$-numbers.}, we employ the identity $[AB, C] = A\{B, C\} - \{A, C\}B$. Using the equal-time anticommutation relations $\{\Psi_\beta(y), \Psi_\gamma(x)\} = 0$ and $\{\Psi_\alpha(x), \Psi_\beta^\dagger(y)\} = \delta_{\alpha\beta} \delta^{(3)}(\vec{x}-\vec{y})$, the commutator simplifies to
\begin{equation}
    [\Psi_\alpha^\dagger(y)\Psi_\beta(y), \Psi_\gamma(x)] = \Psi_\alpha^\dagger(y)\{\Psi_\beta(y), \Psi_\gamma(x)\} - \{\Psi_\alpha^\dagger(y), \Psi_\gamma(x)\}\Psi_\beta(y) = -\delta_{\alpha\gamma} \delta^3(y - x)\Psi_\beta(y).
\end{equation}
Substituting this last result into \eqref{comutador}, we obtain
\begin{equation}
    [iG, \Psi_\gamma(x)] = i\int d^3y \, H_{\alpha\beta}(y)\left(-\delta_{\alpha\gamma} \delta^3(y - x)\Psi_\beta(y)\right) = -H_{\gamma\beta}(x)\Psi_\beta(x) = -(H(x)\Psi(x))_\gamma.
\end{equation}
To evaluate higher-order terms in the expansion, we compute the quadratic contribution, obtaining
\begin{equation}
[iG, -iH \Psi] = -iH [iG, \Psi] = -iH (-iH \Psi) = (-iH)^2 \Psi
\end{equation}
and using this result in \eqref{2vai}, we have 
\begin{equation}
\bar{T} \Psi(x) \bar{T}^{\dagger} = \left( 1 + (-iH) + \frac{1}{2!} (-iH)^2 + \dots \right) \Psi(x)= \exp(-iH(x)) \Psi(x).
\end{equation}
Under this simplified framework, we select the matrix $H$ such that
\begin{equation}
\exp(-iH(x)) = \mathrm{T}(x) = \begin{pmatrix} 0 & 1 \\ \lambda(x) & 0 \end{pmatrix}.
\end{equation}
Since $\mathrm{T}$ is unitary, the existence of the Hermitian matrix $H$ is guaranteed. With this result, we obtain the explicit form of the symmetry operator acting on the doublet space as
\begin{equation}
    \bar{T}_{\Psi} = \exp\left(i \int d^3x \, \Psi^\dagger(x) H(x) \Psi(x)\right),
\end{equation} where $H$ can be expressed, in the usual $M_{2\times 2}(\mathbb{C})$, by 
\begin{equation}
H = -\left( \frac{\theta(x) + \pi}{2} \right) \mathbb{1}_{2\times 2} + \frac{\pi}{2} \left( \sigma_x \cos \frac{\theta(x)}{2} + \sigma_y \sin \frac{\theta(x)}{2}\right),\label{aga}
\end{equation} for instance. 

As an important remark, we observe that the matrix $H(x)$ is not single valued on $V_3$: under $\theta\to\theta+2\pi$ it changes by a nonzero matrix, even though $\mathrm{T}(x)$, whose eigenvalues are $\pm\lambda^{1/2}$, is single valued. The generator $G$ therefore requires a cut $\Sigma$ (cf.\ the footnote on $\lambda_\alpha$) and a choice of branch. The operator $\bar{T}_\Psi$ itself, however, does not depend on this choice. Since $G$ is (ultra) local, $\bar{T}_\Psi$ depends on $H(x)$ only through $e^{iH(x)}=\mathrm{T}^{\dagger}(x)$, which is single valued and smooth on $V_3$. Equivalently, $\bar{T}_\Psi=\Gamma(\mathrm{T}^{\dagger})$
is the second quantization of the one-particle unitary given by multiplication
with $\mathrm{T}^{\dagger}(x)$ on $L^{2} (V_3)\otimes\mathbb{C}^{2}\otimes\mathbb{C}^{2}$, which requires no logarithm. 
Here $\Psi(x)$ is the nonrelativistic electron field, which contains only
annihilation operators: the smeared field $a(f)=\int d^3x\,\overline{f(x)}\cdot\Psi(x)$ is antilinear in the one-particle wave function $f$ and $\Gamma(U)a(f)\Gamma(U)^{-1}=a(Uf)$. This is why $\Gamma(\mathrm{T}^{\dagger})$, and not $\Gamma(\mathrm{T})$, implements $\Psi\mapsto\mathrm{T}\Psi$. Finally, it satisfies $\bar{T}_\Psi\Psi(x)\bar{T}_\Psi^{-1}=\mathrm{T}(x)\Psi(x)$ and is well defined on the Fock space of the nonrelativistic electrons.

We now consider the symmetry operator acting on the magnetic vector potential $\vec{A}$. As in the previous case, our objective is to construct a unitary operator $\bar{T}_A$ that implements the transformation $\bar{T}_A A_{k}(x) \bar{T}_A^{-1} = T(A_{k}(x)) = A_{k}(x) - \mathfrak{B}_{k}(x)$. To this end, invoking Hadamard's lemma once again yields commutators between the field and the generator of the transformation. Assuming a transformation that results in an additive shift $\mathfrak{B}_{k}(x)$, we propose the following Ansatz:
\begin{equation}
\bar{T}_A = \exp \left( -i \int d^3 x \, \mathfrak{B}_{j}(x) \pi^{j}(x) \right).
\label{55}
\end{equation}
Here, $\pi^{j}(x)$ represents the canonical momentum conjugate to the vector potential $A_{j}$\footnote{A gauge-fixing choice is assumed, such as the temporal (or Weyl) gauge $A_0 = 0$, in order to avoid the primary constraint $\pi^0 = 0$ and ensure the validity of the standard commutation relations for the physical degrees of freedom.}, defined by $\pi^{j} = \partial \mathcal{L} / \partial (\partial_0 A_{j})$. %The relevant Lagrangian density is given by
%\begin{equation}\mathcal{L} = -\frac{1}{4} F_{\mu\nu} F^{\mu\nu},
%\end{equation}
%where $F_{\mu\nu}$ is defined by:\begin{equation}F_{\mu\nu} = \partial_\mu A_\nu - \partial_\nu A_\mu.
%\end{equation}
As in Eq. \eqref{2vai}, we compute the action of the transformation operator in Eq. \eqref{55} to first order in the series expansion to get
\begin{equation}
\bar{T}_A A_{k}(y) \bar{T}_A^{-1} = A_{k}(y) - i \int d^3 x \, \mathfrak{B}_{j}(x) [\pi^{j}(x), A_{k}(y)].
\end{equation}
Using $[\pi^{j}(x), A_{k}(y)] = -i \delta^{j}_{k} \delta^3(x-y)$, we have
\begin{equation}
\bar{T}_A A_{k}(y) \bar{T}_A^{-1}= A_{k}(y) - i \int d^3 x \, \mathfrak{B}_{j}(x) (-i \delta^{j}_{k} \delta^3(x-y)).
\end{equation}
Thus, we are left with
\begin{equation}
\bar{T}_A A_{k}(y) \bar{T}_A^{-1}= A_{k}(y) - \mathfrak{B}_{k}(y)
\end{equation}
Since the term generated by the commutator is a $c$-number, the expansion truncates at first order. Moreover, since $\mathfrak B_{j}=\partial_{j}\theta/2q$ has circulation $h/2q$ around the hole, $\bar T_A$ inserts a flux tube of that strength through the hole. If the gauge field is quantized in all of space, this configuration has infinite energy unless regularized and $\bar T_A$ should be read as the automorphism $A_k\mapsto A_k-\mathfrak{B}_k$ of the field algebra, which is all that is used here. Given that equal-time commutators between the fermionic and bosonic sectors vanish, the full symmetry operator is given by the product of both sectors
\begin{equation}
\bar{T} = \bar{T}_{\Psi} \bar{T}_{A} = \exp\left( i \int d^3x \, \Psi^\dagger(x) H(x) \Psi(x) \right) \exp\left( -i \int d^3y \, \mathfrak{B}_{j}(y) \pi^{j}(y) \right),
\label{eq:T_total}
\end{equation}
which implements the symmetry at the quantum level, up to the regularization of the gauge sector discussed above.

\section{Final Remarks}

We have shown how to obtain magnetic flux quantization in a superconducting ring using the exotic Pauli formalism and we explicitly constructed the symmetry operators acting on the Hilbert space, thereby complementing and further developing the application of exotic spinors to superconductivity. The approach pursued here was conservative, in the sense that the Pauli equation was used to investigate nonrelativistic phenomena. A symmetry between exotic and standard spinors can be implemented for the set of relevant observables, driving the flux quantization. Contrasting this formulation with its high-energy counterpart, we find that the symmetry is exact in the Pauli case as well: once the relative signs of the spatial and temporal topological terms are fixed by the covariant exotic Dirac equation, $T^{2}$ is a bona fide transformation and neither a truncation of the exotic operator nor the condition $\partial_{t}\theta=0$ is needed. The only nonrelativistic approximation is the usual reduction of the Dirac equation to the Pauli equation (dropping $O(1/m^{2})$ terms), which is independent of the topology. As for the flux quantization itself, the symmetry yields the periodicity of the spectrum in the flux, with period $h/2q$ fixed by $T$, while the position of the minima rests on the Meissner effect, as in Ref.~\cite{petry}.

Finally, we highlight a connection between Pauli and Dirac exotic spinors in the context of the superconducting ring. Denoting a Dirac spinor by $\psi_D$ and its exotic counterpart by $\tilde{\psi}_D$, it was shown in Ref.~\cite{petry} that the $T$ symmetry acts as $T(\tilde{\psi}_D) = \lambda \psi_D$, the very same phase as in the Pauli case: in both cases $T^{2}$ is a transformation by that phase. 

\section*{Declaration of generative AI and AI-assisted technologies in the manuscript preparation process}

During the preparation of this work, the authors used Claude IA for improving the language and readability of the manuscript and assisting with the \LaTeX{} editing. After using this tool/service, the authors reviewed and edited the content as needed and took full responsibility for the content of the article.

\section*{Acknowledgments}

JMHS thanks CNPq (Grant No. 307641/2022-8) for financial support. RTC thanks CNPq (Grant No. 401567/2023-0), for financial support.

\end{document}